\documentstyle[prd,aps,epsfig,preprint]{revtex}
\begin{document}

\draft

\title{\bf Conservative Force Fields in Nonextensive Kinetic Theory}

\author{J.A.S. Lima\footnote {limajas@dfte.ufrn.br},
J.R.Bezerra\footnote{zero@dfte.ufrn.br}, and R.Silva\footnote
{raimundo@dfte.ufrn.br}}

\smallskip
\address{~\\Universidade
Federal do Rio Grande do Norte,
\\Departamento de F\'{\i}sica,
Caixa Postal 1641, \\59072-970 Natal, RN, Brazil}

\date{\today}
\maketitle
\begin{abstract}
We investigate the nonextensive $q$-distribution function for a
gas in the presence of an external field of force possessing a
potential $U({\bf r})$. In the case of a dilute gas, we show that
the power law distribution including the potential energy factor
term can rigorously be deduced based on kinetic theoretical
arguments. This result is significant as a preliminary to the
discussion of long range interactions according to nonextensive
thermostatistics and the underlying kinetic theory. As an
application, the historical problem of the unbounded isothermal
planetary atmospheres is rediscussed. It is found that the maximum
height for the equilibrium atmosphere is $z_{max} =
K_{B}T/mg(1-q)$. In the extensive limit, the exponential Boltzmann
factor is recovered and the length of the atmosphere becomes
infinite. \newline {\it Keywords}: Nonextensive statistics;
Kinetic theory; Power laws
\end{abstract}

\newpage
In gas dynamics it is very important to investigate how the
molecular motion is modified by force-fields different from those
exerted by the containing vessel or even by the other particles of
the gas. Particular examples are ions in an external magnetic
field and a gas in the earth's gravitational field\cite{Huang,KT}

As widely known, a classical gas under steady state conditions and
immersed in a conservative force field, ${{\bf F}}= -\nabla U({\bf
r})$, is described by a distribution function that differs from
the Maxwellian velocity distribution by an extra exponential
factor involving the potential energy. In this case, the total
equilibrium distribution function reads
\begin{equation}\label{e1}
f({\bf r},v)=n_o\left({m\over 2\pi k_B T} \right)^{3/2}\exp
\left(-{{1\over 2}m v^2 +U({\bf r})\over k_B T}\right),
\end{equation}
where m is the mass of the particles, $T$ is the temperature and
$n_0$ is the particle number density in the absence of the
external force field. In addition, since this distribution
function is normalized, it is easy to see that the number density
is given by
\begin{equation}\label{e2}
n({\bf r})=n_0 \exp \left[-{{U({\bf r}) \over k_B T}}\right],
\end{equation}
where the factor, $\exp[-{U({\bf r})/k_B T}]$, which is
responsible for the inhomogeneity of $f({\bf r},v)$, is usually
called the Boltzmann factor. Expression ($\ref{e1}$) follows
naturally from an integration of the collisionless Boltzmann's
equation
\begin{equation}\label{e3}
\frac{\partial{\it f}}{\partial{\it t}} + {\bf
v}\cdot\frac{\partial{\it f}}{\partial{\bf r}} + \frac{\bf
F}{m}\cdot\frac{\partial{\it f}}{\partial{\bf v}} = 0,
\end{equation}
when stationary conditions are adopted along with the assumption
that the total distribution can be factored
\begin{equation}\label{e4}
f({\bf r},v)=f_0(v)\chi({\bf r}),
\end{equation}
where $f_0(v)$ represents the Maxwell equilibrium distribution
function, and $\chi({\bf r})$ is a scalar function of ${\bf r}$.
As one may show, after a simple normalization, the resulting
expression for $\chi({\bf r})$ is exactly the Boltzmann factor for
the potential energy of the external field, namely:

\begin{equation}\label{e2a}
\chi({\bf r})=\exp \left[-{{U({\bf r}) \over k_B T}}\right],
\end{equation}
and combining this result with equation (\ref{e4}) the Boltzmann
stationary distribution (\ref{e1}) is readily obtained.

On the other hand, recent efforts on the kinetic foundations of
the $q$-nonextensive statistics proposed by
Tsallis\cite{Tsallis,C99} lead to an equilibrium velocity
distribution of the form \cite{SPL98,LSP01}
\begin{equation}\label{e5}
f_0(v) = B_q \left[1 - (1-q){mv^2\over 2k_B T}\right]^{1 \over
{1-q}}.
\end{equation}
In this expression the $q$-parameter quantifies the nonadditivity
property of the associated gas entropy whose main effect at the
level of the distribution function is to replace the standard
Gaussian form by a power law. The quantity $B_q$ is a
$q$-dependent normalization constant whose expression for $1/3 < q
\leq 1$ is given by
\begin{equation}\label{e6}
B_q = n(1-q)^{1/2}\frac{5-3q}{2}\frac{3-q}{2}\frac{\Gamma
(\frac{1}{2}+{1\over 1-q})}{\Gamma({1\over
1-q})}\left(\frac{m}{2\pi k_BT}\right)^{3/2},
\end{equation}
which reduces to the Maxwellian result in the limit $q=1$. As
explained in Ref.\cite{SPL98}, the above power law distribution
can be deduced from two simple requirements: (i) isotropy of the
velocity space, and (ii) a suitable nonextensive generalization of
the Maxwell factorizability condition\cite{M1860}, or
equivalently, the assumption that $F(v)=f(v_x)f(v_y)f(v_z)$. For
$q > 0$, the above distribution function satisfies a generalized
H-theorem, and its reverse has also been proved, that is, the
collisional equilibrium is given by the Tsallis' q-nonextensive
velocity distribution\cite{LSP01}.

In the last few years, several applications of this equilibrium
power law velocity distribution have been done in many disparate
branches of physics [8-14]. It is worth notice that the widely
spread belief that Tsallis equilibrium distribution cannot be
applied to Hamiltonian systems seems to be a profound
misunderstanding on the foundations of statistical mechanics and
kinetic theory. Indeed, the BG canonical emsemble approach is
valid only for sufficiently short-range interactions. It fails
when gravitational or unscreened Coulombian fields are present,
that is, to the class of systems where the usually assumed entropy
additivity postulate is not valid. In other words, when the
dynamics plays a nontrivial role, the system does not relaxes to
the BG distribution, but they evolve to a non-Gaussian velocity
distribution which can be fitted by the monoparametric class of
Tsallis' nonextensive statistics (for a more detailed discussion
see \cite{L01,CT02}).

In this letter we investigate how the potential energy term can
rigorously be introduced in the $q$-distribution with basis on the
kinetic approach. More precisely, it is shown that an analytical
expression for the equilibrium distribution of a dilute gas under
the action of a conservative field may also be calculated as a
stationary solution of the collisionless Boltzmann equation. This
result is significant as a preliminary to the discussion of long
range interactions according to nonextensive thermostatistics and
the underlying kinetic theory. As an illustration we discuss the
case of planetary atmospheres. It is found that the problem of an
infinite isothermal atmosphere is naturally solved in this
extended framework. For a given value of $q<1$, the extension of
the atmosphere becomes finite and its length is uniquely
determined by the temperature and mass of the molecular
components.

Let us now consider a spatially inhomogeneous dilute gas supposed
in equilibrium at temperature $T$. It is immersed in a
conservative external field in such a way that $f(r,v)d^3v d^3r$
is the number of particles with velocity lying within a volume
element $d^3 v$ about ${\bf v}$ and positions lying within a
volume element $d^3 r$ around ${\bf r}$. In this case, we see from
(\ref{e6}) that the stationary Boltzmann equation can be rewritten
as
\begin{equation}\label{e7}
{\bf v}.\nabla_r f - \frac{1}{m}\nabla_r U.\nabla_v f = 0.
\end{equation}

In order to introduce the nonextensive effects we first recall
that the factorizability condition cannot be applied in this
extended framework. This means that the standard starting
assumption (see equation (\ref{e4})) must be somewhat modified. In
the spirit of the $q$-nonextensive formalism, a consistent
$q$-generalization of (\ref{e4}) is
\begin{equation}\label{e8}
f({\bf r} ,v) = B_q e_q\left[ \ln_q\left(\frac {f_0}{B_q}\right) +
\ln_q \chi({\bf r})\right],
\end{equation}
where $B_q$ is the constant expression given by (\ref{e6}) which
has been introduced for mathematical convenience, and the
functions $q$-exp and $q$-log, $e_q(f)$, $\ln_q(f)$, are defined
by
\begin{equation}\label{e9}
e_q(f)=[1+(1-q)f]^{1/1-q},
\end{equation}
\begin{equation}\label{e10}
\ln_q f= {f^{1-q}-1\over 1-q}.
\end{equation}
Note that in the limit ${q \rightarrow 1}$ the above identities
reproduce the usual properties of the exponential and logarithm
functions. In addition, $e_q(\ln_q f) = \ln_q (e_q(f)) = f$, and
as should be expected from (\ref{e8}), the factored decomposition
(\ref{e4}) is readily recovered in the extensive limit. In what
follows, the properties of $q$-exponential and $q$-log
differentiation
\begin{eqnarray}\label{e11}
{d\ln_q f \over dx}= f^{-q}{df \over dx}, \\ {d e_q(f) \over
dx}=e^q_q(f){df \over dx},
\end{eqnarray}
will also be extensively used. In particular, for the total
$q$-distribution (\ref{e8}), we obtain
\begin{equation}\label{e12}
\nabla_v f = -e^q_q(x)\frac{m\bf v}{k_BT}.
\end{equation}
Now, substituting $\nabla_r f$ and $\nabla_v f$ into the
stationary Boltzmann equation (\ref{e7}), and performing the
elementary calculations one obtains
\begin{equation}\label{e13}
\chi^{-q}\nabla \chi.d{\bf r}= - \frac{1}{k_BT}\nabla U.d{\bf r}
\end{equation}
the solution of which is
\begin{equation}\label{e14}
\chi({\bf r})= e_q \left( - \frac{U({\bf r})}{k_BT} + C \right),
\end{equation}
where $C$ is an arbitrary constant.

Now, inserting (\ref{e14}) into (\ref{e8}), and integrating the
result in the velocity space it follows that
\begin{equation}\label{e15}
\int B_q e_q\left[\ln_q\left(\frac {f_0}{B_q}\right) -
\frac{U}{k_BT} + C \right] d^3v = n({\bf r}).
\end{equation}
Finally, by substituting the expression of $f_0$ and considering a
region where $U({\bf r}) =0$, one finds
\begin{equation}\label{e16}
B_q\int e_q\left(- \frac{m{\bf v}^2}{2k_BT} + C \right) d^3v =
n_0,
\end{equation}
and from normalization condition, $n_0 = \int f_0(v) d^3v$, it
follows that the unique allowed value for the integration constant
is $C=0$. Consequently, (\ref{e14}) becomes
\begin{equation}\label{e17}
\chi ({\bf r})= e_q\left[- \frac{U({\bf r})}{k_BT}\right],
\end{equation}
which is the $q$-generalized Boltzmann factor.

Finally, inserting this result into (\ref{e8}), we obtain the
complete $q$-distribution function in the presence of an external
field
\begin{equation}\label{eq18}
f({\bf r},v) = B_q\left[1 -(1-q)\left(\frac{m{\bf v}^2}{2k_BT} +
\frac{U({\bf r})}{k_BT}\right)\right]^{1/(1-q)}\equiv B_q
e_q(-E/k_B T),
\end{equation}
where $E$ is the total energy of the particles. It thus follows
that a generalized $q$-exp factor for Tsallis' nonextensive
thermostatistics can exactly be deduced if the standard approach
is slightly modified. The particle number density, $n({\bf r}) =
\int f ({\bf r}, {\bf v})d^3v$, also becomes a function of the
position given by

\begin{equation}\label{eq19}
n({\bf r}) = n_0 \left[1 -(1-q)\frac{U({\bf
r})}{k_BT}\right]^{(5-3q)/2(1-q)} \equiv
n_0[e_q(-U/k_BT)]^{\frac{5-3q}{2}} \quad,
\end{equation}
and as should be expected, in the extensive limit ($q \rightarrow
1$), the standard exponential Maxwellian expressions for $f({\bf
r},v)$ and $n({\bf r}$) are readily recovered.

At this point it is interesting to analyze some application of the
above result. We recall that the most popular problem of a gas in
a force-field is the planetary atmosphere. In the standard
simplified treatment, the temperature is uniform and the
tree-dimensional motion occurs under the action of a constant
gravitational field ${\bf g}$ along the z-direction. Thus, each
molecule has a potential energy, $U(z) = mgz$, so that the
concentration of particles, or equivalently, the gas density
($\rho = nm$) is given by the familiar barometric formula (see Eq.
(\ref{e2})).

\begin{equation}\label{eq20}
\rho(z)= \rho_0\exp\left[{-\frac{mgz}{k_BT}}\right] \quad,
\end{equation}
with the pressure obeying a similar exponential law in virtue of
the standard equation of state for a perfect gas. The simplest
conclusion based on the above expression is that an ideal
isothermal atmosphere would extend to indefinite distances from
the planet surface. This result disagree completely with the
existing observations even by considering that the upper-most part
of the earth atmosphere is nearly in thermodynamic equilibrium.
Since there is no a natural boundary in this Boltzmann like
picture, a possible solution to this problem is to assume that the
atmospheres of the planets are actually finite, although being
only in a pseudo-equilibrium state with the particles continuously
streaming off into space at a very slow rate (evaporation).

How does this basic prediction is modified when an equilibrium
$q$-distribution is assumed? The nonextensive answer to this
question is immediate. The new distribution itself is given by
(see (\ref{eq18}))

\begin{equation}\label{eq21}
f({\bf r},v) = B_q\left[1 -(1-q)\left( \frac{m{\bf v}^2}{2k_BT} +
\frac{mgz}{k_BT}\right)\right]^{1/(1-q)} \quad,
\end{equation}
while from (\ref{eq19}), the matter density assumes the form

\begin{equation}
\label{eq22} \rho(z) = \rho_0\left[1
-(1-q)\frac{mgz}{k_BT}\right]^{(5-3q)/2(1-q)}.
\end{equation}
For comparison, in Fig. 1 we plot the gas density as a function of
$z$ for some values of the $q$-parameter including the
Boltzmannian case. Note that the standard exponential curve
($q=1$) is replaced by the characteristic power law behavior of
Tsallis' nonextensive framework. In a more realistic treatment, as
in the case of mixed atmospheres, the above distribution
(\ref{eq21})-(\ref{eq23}) may separately be applied to each
component. As in the Boltzmann case, heavier gases are more
concentrated near the surface of the earth than the lighter ones.
In particular, this means that at greater heights the fraction of
oxygen is smaller than helium as compared to the corresponding
fractions at the surface. Such a result remains true as long as
the maximum height allowed for a given component is not attained.
In fact, for values of $q$ smaller than unity, the positiviness of
the argument appearing in the power law implies that
($\ref{eq22}$) exhibits a ``thermal cut-off'' to the allowed
values of the $z$-coordinate which defines the length of the
atmosphere. It thus follows that the maximum height of the
atmosphere (or more generally, the length of any isothermal layer)
is

\begin{equation}\label{eq23}
z_{max} = {k_{B}T \over mg(1-q)} \quad.
\end{equation}
We see that for given values of the temperature and $q<1$, the
maximum height associated to a specific component is inversely
proportional to its molecular mass. In table 1, we display the
predicted values of $z_{max}$ for a temperature $T = 298,15 K$ and
two different kind of gases (Oxygen and Hydrogen). In figure 1,
one may see two different plots showing the finiteness of the
atmosphere for values of $q$ smaller than unity. Note that
$z_{max}$ increases without limit when $q \rightarrow 1$
recovering the Maxwellian result. As one may check, the predicted
atmosphere is infinite for all values of $q \geq 1$ since the
matter density, as given by equation (\ref{eq22}), does not vanish
at any finite height.

The above solution to the isothermal atmosphere with infinite
length resembles a similar solution proposed in galactic dynamics
to the collisionless stellar isothermal sphere. In the latter
case, the Maxwell-Boltzmann distribution also predicts an infinite
system whose integrated density profile yields an infinite total
galactic mass. This theoretical problem was also solved with help
of the nonextensive power law distribution\cite{plastino}. In the
case of planetary atmosphere, the nonextensive prediction of a
finite length, as well as that heavier components are associated
to a smaller maximum height could be tested by experiments using a
mass spectrograph in atmospheric balloons.

\begin{center}
\begin{table*}
\centering
\begin{center}
\caption{Limits to the height of the atmosphere}
\end{center} \begin{tabular}{rll}
\hline \multicolumn{1}{c}{Gas ($T=298,15 \ K$})&
\multicolumn{1}{c}{$q$}& \multicolumn{1}{c}{$z_{max}$} \ ({\rm
Km})\\ \hline Hydrogen:..............& 1.0 & $\infty$ \\ & 0.9 &
1239
\\
 & 0.8 & 619.5 \\
& 0.7 & 413
\\ Oxygen:..................& 1.0 & $\infty$ \\ & 0.9
& 77.4
\\
 & 0.8 & 33.7 \\
& 0.7 & 25.8\\ \hline
\end{tabular}
\end{table*}
\end{center}

\begin{figure}
\vspace{.2in}
\centerline{\epsfig{figure=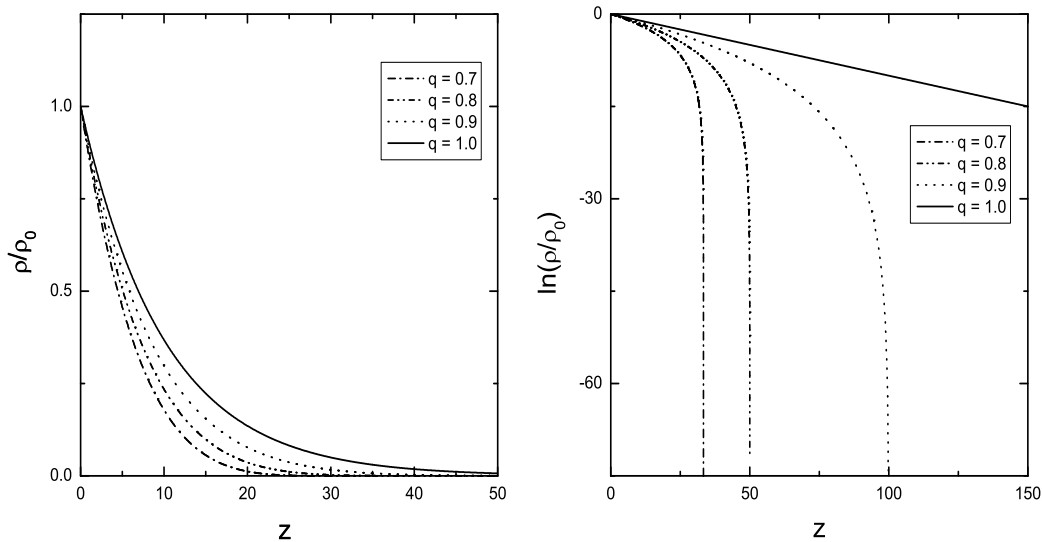,width=6.0truein,height=3.5truein}
\hskip 0.1in} \caption{Maximum height of the isothermal
atmosphere. The gas density as a function of $z$ for some selected
values of the $q$ parameter. In the left panel, the solid line
($q=1$) is the exponential curve for an infinite atmosphere
predicted by the standard Boltzmann distribution. For values of
$q<1$ all curves decrease rapidly toward the maximum value of $z$
which is a physical cut-off present in the power law behavior. The
right panel displays the logarithm of the gas density versus $z$
making more clear the existence of this cut-off. Note that the
standard curve ($q=1$) is a straight line going to infinite.
However, for values of $q<1$, there is a knee causing an
intersection of the curves at finite values of the z
axis.}\label{gra2}
\end{figure}

As a conclusion, we call attention for some simple physical
situations where the power law distribution (\ref{eq22}) can play
an interesting role. The first example is the question concerning
the escape of molecules to infinity. This phenomenon seems to be
more easily understood in this extended framework since the power
law distribution naturally defines a limiting boundary frontier
where the atmosphere terminates (see Table 1 and Fig. 1). We
recall that in the standard treatment, the rate of loss to
infinity is based on the assumption that the upper part of the
atmosphere extends in isothermal equilibrium only until a certain
height $h$. In other words, unlike in the $q$-nonextensive
framework, there is a cut-off introduced by hand in the infinity
isothermal atmosphere predicted by the Boltzmann approach. A more
quantitative analysis of evaporation taking into account the earth
curvature will be presented in a forthcoming communication.

Other examples of forces, like the ones involving giroscopic terms
may also be trivially added to the power law distribution
(\ref{eq22}). For rotating frames with constant angular velocity,
for instance, the effect is just to add the term
$-1/2m\omega^{2}l^{2}$ to the potential energy, where $l$ is the
perpendicular distance from the axis of rotation. As usual, this
term simulates a slight change in the potential energy due to
gravity\cite{LL}.

Finally, we also remark that for colloidal particles in solutions,
the gravity acts on the particles which are buoyed up by the
liquid in which they are suspended. As a consequence, instead of
the weight $mg$ the net force is $mg(\rho_p - \rho_l)/\rho_p$,
where $\rho_p$ and $\rho_l$ are the particles and liquid
densities, respectively. In this case, the resulting nonextensive
expression for $n(z)$ could experimentally be tested by repeating
the experiments with colloidal suspensions as originally done by
Perrin (see \cite{PE20} and references there in). Naturally, the
inclusion of other physical effects leading to independent probes
of the q-statistics take even greater interest.

\vspace{1.0cm}

\noindent {\bf Acknowledgments:} The authors are grateful to
Odylio Aguiar for helpful discussions. This work was supported by
Pronex/FINEP (No. 41.96.0908.00), Conselho Nacional de
Desenvolvimento Cient\'{\i}fico e Tecnol\'{o}gico - CNPq and CAPES
(Brazilian Research Agencies).

\end{document}